\documentclass[aps,pra,showpacs,twocolumn,amsmath,amssymb,superscriptaddress, footinbib]{revtex4}

\usepackage[english]{babel}

\usepackage{latexsym}
\usepackage{graphicx}
\usepackage{subfigure}
\usepackage{epsfig}
\usepackage{amsfonts}
\usepackage{amssymb}
\usepackage{amsmath}
\usepackage{bm, bbm}

\usepackage{lipsum} 
\usepackage[draft]{todonotes}   

\begin{document}

\title{Phases of $d$-orbital bosons in optical lattices}

\author{Fernanda Pinheiro}
\affiliation{Department of Physics,
Stockholm University, AlbaNova University Center, 106 91 Stockholm,
Sweden}
\affiliation{NORDITA, KTH Royal Institute of Technology and Stockholm University, Se-106 91 Stockholm, Sweden}
\affiliation{Institut f\"ur Theoretische Physik, Universit\"at zu K\"oln,
De-50937 K\"oln, Germany}
\author{Jani-Petri Matrikainen}
\affiliation{COMP Center of Excellence, Department of Applied Physics, Aalto University, Fi-00076, Aalto, Finland}
\author{Jonas Larson}
\affiliation{Department of Physics,
Stockholm University, AlbaNova University Center, 106 91 Stockholm,
Sweden}
\affiliation{Institut f\"ur Theoretische Physik, Universit\"at zu K\"oln,
De-50937 K\"oln, Germany}

\date{\today}

\begin{abstract}
We explore the properties of bosonic atoms loaded into the $d$-bands of an isotropic square optical lattice. Following the recent experimental success reported in [Y. Zhai {\it et al.}, Phys. Rev. A {\bf 87}, 063638 (2013)], in which populating $d$ bands with a 99$\%$ fidelity was demonstrated, we present a theoretical study of the possible phases that can appear in this system. Using the Gutzwiller ansatz for the three $d$ band orbitals we map the boundaries of the Mott insulating phases. For not too large occupation, two of the orbitals are predominantly occupied, while the third, of a slightly higher energy, remains almost unpopulated. In this regime, in the superfluid phase we find the formation of a vortex lattice, where the vortices come in vortex/anti-vortex pairs with two pairs locked to every site. Due to the orientation of the vortices time-reversal symmetry is spontaneously broken. This state also breaks a discrete $\mathbb{Z}_2$-symmetry. We further derive an effective spin-1/2 model that describe the relevant physics of the lowest Mott-phase with unit filling. We argue that the corresponding two dimensional phase diagram should be rich with several different phases. We also explain how to generate antisymmetric spin interactions that can give rise to novel effects like spin canting. \end{abstract}

\maketitle

\section{Introduction} 
Interest in systems of cold atoms in optical lattices has greatly increased during the last decade~\cite{rev} partly because of their
versatility as simulators of quantum systems. More precisely, the flexibility, control, and cleanness of these systems have led to numerous realisations of paradigm many-body models of condensed matter physics. The field was
greatly stimulated
with experimental explorations of the {\it Mott insulator-superfluid transition} of the {\it Bose-Hubbard} (BH) model~\cite{bloch} following the theoretical proposal of Ref.~\cite{BH}. Today the list of achievements is long, but to mention just a few: single-site resolved detection~\cite{singlesitedet}, simulation of magnetism and spin models~\cite{qmagn}, realising {\it synthetic magnetic fields}~\cite{syngauge} and {\it topological states}~\cite{topo}, demonstration of a {\it fermion Mott insulator}~\cite{FH}, as well as various dynamical studies like equilibration and {\it Lieb-Robinson} spread of correlations~\cite{dynamics}. Common to all the above examples is that the essential physics appear on the lowest energy band, the $s$ band. In particular, for spinless particles on the $s$ band, the onsite atomic states are not degenerate. We know, however, that many interesting phenomena in condensed matter theories have their origin 
in the degeneracies of onsite states/orbitals. For example, understanding 'orbital physics'~\cite{orbital} is essential for giving a full description of superfluid properties of $^3$He~\cite{he3} and the metal-insulating transitions in {\it metal oxides}~\cite{metal}. Spurred by this, in recent years the first steps towards controlled studies of orbital physics with cold atoms have been taken~\cite{dance}.

Degenerate orbital atomic states appear naturally on excited bands of optical lattices, and it was predicted that bosonic atoms loaded into the first excited, the $p$ bands, of a two dimensional (2D) square optical lattice give rise to interesting physics beyond that found on the lowest band~\cite{isacsson}. In particular, the superfluid order parameter at zero temperature is complex valued and time-reversal symmetry is spontaneously broken~\cite{isacsson,liu1,larson2}. Also the insulating phases provide rich physics with the possibilities to study exotic quantum magnetic phases~\cite{mottp,fernanda2}. In different lattice geometries, the system may display 'stripped phases'~\cite{pbandphases}, and in 3D the state of bosons occupying an isotropic cubic lattice become frustrated both in the insulating as well as in the {\it superfluid phase}~\cite{frustration,fernanda3}. Fermionic systems in the $p$ band are also known to feature rich physics, with a plethora of different phases~\cite{pbandfermion}. 

Experimentally, populating the $p$ bands with bosonic atoms was first realised by accelerating the lattice potential in such a way that the system traversed through a {\it Landau-Zener transition} non-adiabatically~\cite{latticeacc}. However, a more thorough experimental analysis of the coherence and life-time of $p$-band bosons was performed by M\"{u}ller {\it et al.}~\cite{pbandbloch} using a two-laser {\it Raman} loading technique from an insulating $s$-band state. In particular, it was found that the atoms on the $p$ band relax very rapidly (on the order of a few tunneling times), and that the life-time is relatively long in comparison to the tunneling time. This enables coherence to be established
from the initial Mott insulator and the possibility to detect dynamical processes. More recently, a 2D superlattice was used in order to prepare bosonic atoms in hybridised states composed of $s$ and $p$ orbitals~\cite{pbandhemmerich}. In this experiment, and in a follow up one~\cite{hemmerich2}, clear evidences of a complex order parameter were presented. The same loading method was further benchmarked by loading atoms into $f$-orbital states~\cite{fbandhemmerich}, i.e. the third set of excited bands.

A completely different approach to initialise bosonic atoms in excited bands was proposed and demonstrated by Zhai {\it et al.}~\cite{dband}. There the atoms start in a zero momentum eigenstate (held in a broad harmonic trap) and then a sequence of `on/off pulses' are applied to the atoms, where the `on pulse' consists of a standing wave which couple the zero momentum states to higher momentum states by photon recoils. With this method the atoms were loaded into the $d$ band with a fidelity as high as 99$\%$, which is much better than what was reported for the $p$ bands (around 80$\%$). This technique is, however, more easily applied for atom transfer between bands of even or odd parities, e.g. $s\rightarrow d$ or $p\rightarrow f$ and so on, and is therefore not suitable for preparing $p$-band atoms starting from the ground state. Surprisingly, the properties of $d$-band bosons remain unexplored. Considering the recent experiment ~\cite{dband}, it is therefore of importance to work out the relevant physics of such systems.

Here we analyse the zero temperature phase diagram of $d$-band bosons in an isotropic 2D square lattice. On the $d$ bands we have three orbitals which are generally denoted $d_{x^2}$, $d_{y^2}$, and $d_{xy}$. However, due to the anharmonicity of the single site potential well, degeneracy is broken and
the energy of the $d_{xy}$ orbital is somewhat higher than the other two orbitals which stay degenerate. The Mott insulator boundaries are determined within the {\it Gutzwiller method}. At low atomic densities, the $d_{xy}$-orbitals are only weakly populated and we can therefor reduce the number of orbitals to two, i.e. we can describe the physics with an effective two-orbital BH model. The properties of the superfluid phase are analysed analytically by minimising the onsite energies. Like for the $p$-band superfluid phase, the order parameter is complex but due to the particular shape of the $d$ orbitals a new type of vortex lattice is formed with two vortex/anti-vortex pairs fixed to every site. At higher densities, the general structure of the vortex lattice persists but now also the $d_{xy}$-orbitals becomes populated.

One difference with respect to $p$-band BH models is the lack of a characteristic $\mathbb{Z}_2$-parity symmetry for bosonic atoms in the $d$ band. The lack of this symmetry stems from density-assisted orbital-changing collision terms. Nevertheless, there is another $\mathbb{Z}_2$-symmetry that is spontaneously broken in the superfluid phase. In the Mott insulating phase, we employ perturbation theory to derive an effective spin model, where it is seen that the new interaction terms appear in the form of an external field. In particular, the first insulating phase (the Mott with unit filling) can be described by an $XY\!Z$ {\it Heisenberg model} in an external field. The model is, of course, still supporting the $\mathbb{Z}_2$-symmetry and we expect a rich phase diagram within the insulating phase with the possibility of ferromagnetic phases with broken symmetry. It should be noted that the phase diagram of the 2D $XY\!Z$ model in an external field is not fully known. We also present a brief discussion on how the so called {\it Dzyaloshinsky-Moriya} (DM) interactions (or antisymmetric spin exchange) may appear when the lattice is no longer isotropic but the onsite degeneracy is kept. While not studied in detail, we point out that the insulating phases
with higher fillings are also of great interest. Such phases will represent exotic magnetic states of higher spins or quasi-spins of $SU(N)$ models.

\section{$d$-band bosons}
Throughout we will assume that all atoms reside on the $d$ bands, i.e. the loading is perfect and that any time-scales are short compared to the decay times, driven by intra-atomic collisions, of these meta-stable atoms. The second condition implies that relaxation occurs on a time shorter than a few $\mu$s which is the experimentally measured life-time of $d$-band atoms~\cite{dband}. Relaxation typically happens during few tunneling times~\cite{pbandbloch}, which allows coherence on these excited bands to form before substantial dissipation/decoherence sets in. This has in particular been experimentally demonstrated for bosons loaded in the $p$~\cite{pbandbloch,pbandhemmerich} and the $f$~\cite{fbandhemmerich} bands. 

Before discussing the physics on the $d$ bands, let us first recapitulate the main results for the system of bosons in the $p$ bands of $2D$ square lattices.
On the isotropic lattice, $p$-band bosons support two degenerate orbitals and a very interesting phase diagram. As for the regular $s$-band BH model~\cite{BH,BH1}, the interplay between repulsive interaction and kinetic hopping leads to Mott insulating phases with integer fillings $n_0=0,\,1,\,2,...$ and condensed superfluid phases~\cite{isacsson,larson2}. The orbital degree of freedom implies, however, that both the insulating and the superfluid phases are non-trivial. In the condensate, the superfluid order parameter $\psi(x,y)$ is a complex function representing a condensate with a vortex pinned to every site. The properties of the tunneling of $p$-band atoms in 2D further implies that vortices in neighboring sites have angular momentum quantised in opposite directions (i.e. winding numbers are $\pm 1$)~\cite{isacsson,larson2,liu1}. Thus, there is a `checkerboard' lattice of vortices and the two possible configurations underline the discrete $\mathbb{Z}_2$-parity symmetry of the model, as well as the spontaneous breaking of time-reversal symmetry. The orbital degree of freedom also makes the insulating phases very rich. The $n_0=1$ Mott phase with one atom per site can be mapped onto an effective Heisenberg $XY\!Z$ model~\cite{fernanda2}, whose phase diagram in 1D is qualitatively known~\cite{sela} but not known in 2D. In the spin language of the Heisenberg $XY\!Z$ model, the breaking of the $\mathbb{Z}_2$-parity symmetry represents for example an {\it Ising-like transition}. 

The goal of the present paper is to demonstrate that also the experimentally relevant $d$-band bosonic model hosts very interesting physics with some notable differences from the physics in the $p$ band. A first guess would be that on the $d$ bands, one again encounters a checkerboard of quantised vortices but with winding numbers $\pm2$ instead. This would be the direct generalisation of the results appearing on the $p$ bands. To construct such states all three orbitals need to be populated, but as we will see in the following only two of them are actually
populated due to broken degeneracy. As a result we find a different type 
of vortex lattice. However, before discussing the possible phases on the $d$ band, we give the corresponding many-body Hamiltonian in the next Section, and outline how we will   analyse it within the Gutzwiller mean-field approach.

\subsection{Single particle properties} 

\begin{figure}[h]
\centerline{\includegraphics[width=8cm]{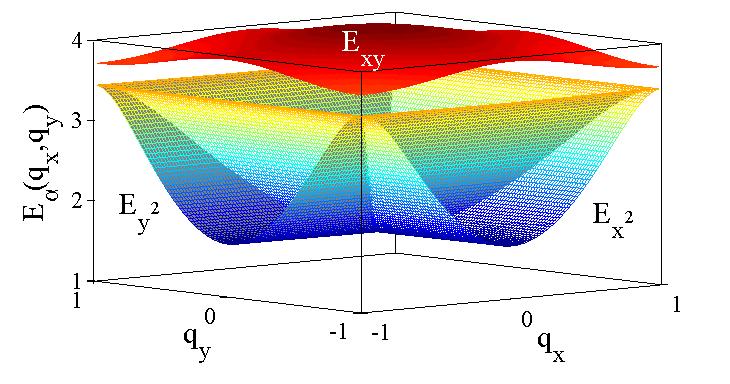}}
\caption{The three $d$-bands; $E_{x^2}(q_x,q_y)$, $E_{y^2}(q_x,q_y)$, and $E_{xy}(q_x,q_y)$ for the potential amplitude $V=20E_r$. The bandwidth gives an estimate of the tunneling strength and the curvature the sign of the tunneling coefficient} \label{fig1}
\end{figure} 

The single particle Hamiltonian for an atom with mass $m$ in an isotropic 2D square optical lattice is~\cite{rev}
\begin{equation}
\hat{H}_\mathrm{sp}'=-\frac{\hbar^2}{2m}\left(\partial_x^2+\partial_y^2\right)+ \tilde{V}\left[\sin^2(kx)+\sin^2(ky)\right].
\end{equation}
For convenience we introduce dimensionless parameters by letting the recoil energy $E_\mathrm{r}=\hbar^2k^2/2m$ set the energy scale and the inverse wave number $k^{-1}$ the characteristic length scale. Thus, in dimensionless variables, $\hat{H}_\mathrm{sp}=\frac{\hat{H}_\mathrm{sp}'}{E_\mathrm{r}}=-\partial_x^2- \partial_y^2+V\left[\sin^2(x)+\sin^2(y)\right]$, with $V=\tilde{V}/E_\mathrm{r}$. The  spectrum $E_\nu(q_x,q_y)$, consisting of bands separated by forbidden gaps, is characterised by three quantum numbers, a discrete {\it band index} $\nu$ and two continuous {\it quasi  momenta} $q_x$ and $q_y$. In scaled dimensions, the first {\it Brillouin zone} is defined by $q_x,\,q_y\in[-1,+1)$. Increasing the potential amplitude $V$ implies that the band gap increases while the widths of the energy bands shrink. The `flatness' of the bands determines the mobility of the atom. Consequently, as $V$ increases the mobility is reduced and the bands become flatter. In the limit of very deep lattice each site mimics a two-dimensional harmonic oscillator and hence the isotropic $2D$ case is characterised by a single $s$ band, two degenerate $p$ bands, three degenerate $d$ bands and so on for the higher excited bands. The similarity with an isotropic $2D$ harmonic oscillator can also be seen from the quantum numbers of the onsite orbitals; if $\vert n_x, n_y\rangle$ represents the eigenstate of the oscillator with quantum numbers $n_x$ and $n_y$ then the ground-state is the `vacuum' $|0,0\rangle$ ($s$-orbital state), the first excited states are $|0,1\rangle$ and $|1,0\rangle$ ($p$-orbital states), and the second excited states are $|0,2\rangle$, $|2,0\rangle$, and $|1,1\rangle$ ($d$-orbital states).  For finite, but large $V$, the separations between the different set of bands ($s,\,p,\,d,...$) may remain much larger than the band widths (at least for the lower bands). However, due to the anharmonicity of the potential wells the bands are no longer equidistant from each other, and even the three $d$ bands may split. As an example, in Fig.~\ref{fig1} we display the three $d$ bands
($E_{x^2}$,$E_{y^2}$, and $E_{xy}$) for a potential of amplitude $V=20$. As we can see, the  $d_{xy}$ band has a higher energy than the other two.

\begin{figure}[h]
\centerline{\includegraphics[width=8cm]{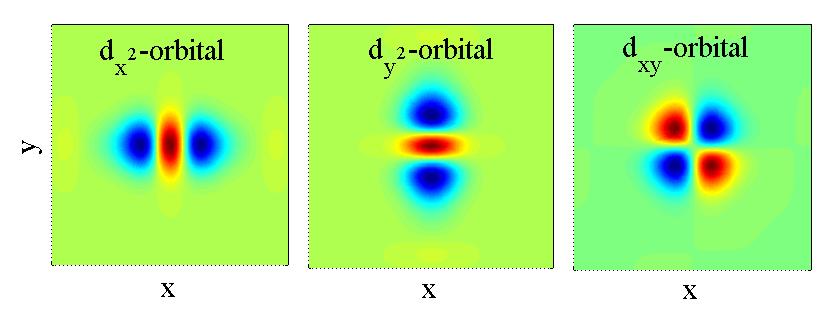}}
\caption{The three $d$-band Wannier orbitals  of an isotropic 2D square lattice. Here, blue/red colour represents negative/positive values. Thus, the $d_{x^2}$ and $d_{y^2}$ orbitals have each two nodes in either the $x$- or the $y$-direction respectively, while the $d_{xy}$-orbit has a single node in both directions. } \label{fig2} 
\end{figure}

The eigenstates of the single particle Hamiltonian $\hat{H}_\mathrm{sp}$ are the {\it Bloch states} $\phi_{\bm \nu {\bf q}}({\bf x})$ with $\bm \nu = (n_x, n_y)$ the band index, ${\bf q}=(q_x,q_y)$ the quasimomentum and ${\bf x}=(x,y)$. Taking the modified (i.e. restrict the integration to the first Brillouin zone) Fourier transform of these, one obtains the {\it Wannier functions} $w_{\bm \nu\mathbf{i}}({\bf x})$ where $\mathbf{i}=(i_x,i_y)$ with $i_x,\,i_y\in\mathcal{N}$ labelling the site. Contrary to the de-localised Bloch functions, the  Wannier functions are localised around the site $\bm i$ and can therefore be normalised in the usual way. In addition, although they are not the eigenfunctions of the single-particle problem, Wannier functions are orthogonal and therefore provide an alternative basis for describing such systems. In the infinitely deep lattice limit, they become simply the harmonic oscillator $\langle{\bf x}|n_x,n_y\rangle$ eigenfunctions localised at each site. For large but finite $V$, the harmonic oscillator eigenfunctions are not exact, but they still preserve the general structure of $w_{\bm \nu\bm i}(\bm x)$. That is, for the $s$ band, where $\nu = (0,0)$, $w_{s{\bf i}}({\bf x})$ is approximately Gaussian. For the $p$ bands, of $\nu = (1, 0)$ and $(0,1)$, the Wannier functions $w_{p_x{\bf i}}({\bf x})$ and $w_{p_y{\bf i}}({\bf x})$ are approximately Gaussian in one direction and have a node in the transverse direction (the subscripts $x$ and $y$ tell the direction of the node). In the $d$ bands, $\nu = (2, 0),\, (1,1)$, and $(0,2)$, the three Wannier functions are shown in Fig.~\ref{fig2}. These are the three atomic $d$-orbital states, and the notation used is borrowed from atomic/chemical physics: The $d_{x^2}$-orbital state has two nodes in the $x$ direction and no node in the $y$ direction. The opposite is true for the $d_{y^2}$ orbital, while the $d_{xy}$ orbital has a single node in both directions. Furthermore, since the square lattice is separable, Wannier functions in 2D can be constructed as a product of 1D Wannier functions. Explicit expressions for the $d$-band Wannier functions are:
\begin{equation}\label{wans}
\begin{array}{l}
w_{d_{x^2}{\bf i}}({\bf x})=w_{s{\bf i}}(y)w_{d{\bf i}}(x),\\ \\
w_{d_{y^2}{\bf i}}({\bf x})=w_{d{\bf i}}(y)w_{s{\bf i}}(x),\\ \\
w_{d_{xy}{\bf i}}({\bf x})=w_{p{\bf i}}(x)w_{p{\bf i}}(y),
\end{array}
\end{equation}
where we have introduced the subscripts $s,\,p,\,d$ to label the Wannier functions of the different bands in 1D. 

Naturally, the shape of the Wannier functions play a crucial role in the dynamics of the system: On a single-particle level, since the $d_\alpha$ orbital ($\alpha=x^2,\,y^2$) is much wider in the direction of the two nodes, tunneling processes parallel to this direction happen with considerably larger amplitude than in the perpendicular one. Thus, a $d_{x^2}$-orbital atom is more mobile in the $x$ than in the $y$ direction. The $d_{xy}$ orbital, however, is equally mobile in both directions. This direction-dependent propagation results in an anisotropy in the problem, that is also present in the $p$-band system~\cite{isacsson,larson2,liu1} (although absent on the $s$-band). This is also reflected as an anisotropy in the single-particle spectrum. As illustrated in Fig.~\ref{fig1}, $E_{x^2}$ has a much larger curvature (inverse effective mass) in the $x$ direction than in the $y$ direction. Moreover, the sign of the curvature is also of importance since it determines whether the effective mass is positive (particle-like) or negative (hole-like), and as we will see below, it also sets the sign of the tunneling coefficient in the extended BH model. Furthermore, on a many-body level the amplitude of scattering processes is determined by overlaps of products of Wannier functions, and therefore their shape strongly influences the possible interaction processes as well as their strengths. 

\subsection{Many-body Hamiltonian}
The Hamiltonian of the full many-body problem describing contact interactions between the neutral atoms is given by 
\begin{equation}\label{many-body_Ham}
\hat{H}_\mathrm{BH} \!=\!\int \!d{\bf x}\!\left[\!\hat \Psi^\dagger({\bf x}) \hat{H}_\mathrm{sp}\hat \Psi({\bf x})  + \!\frac{U_0}{2}\hat \Psi^\dagger({\bf x})\hat \Psi^\dagger({\bf x})\hat \Psi({\bf x})\hat \Psi({\bf x}) \!\right]\!,
\end{equation}
where the coupling constant $U_0 =
4\pi\hbar^2a/m$, with $m$ the atomic mass and $a$ the s-wave scattering length. The field operators $\hat \Psi({\bf x})$ and $\hat \Psi^\dagger({\bf x})$ annihilate and create a particle at the position ${\bf x}$ and obey the bosonic commutation relations $[\hat \Psi({\bf x}),\hat \Psi^\dagger({\bf x}^{'}) ] = \delta({\bf x} -  {\bf x}^{'})$. 

In order to constrain the atoms to the $d$ bands (i.e., impose the {\it single-band approximation}), we expand the field operators in the corresponding Wannier functions,  
\begin{equation}\label{wanexp}
\hat{\Psi}
(\mathbf{x})=\sum_\mathbf{i}\hat{d}_{{x^2}\mathbf{i}}w_{d_{x^2}\mathbf{i}}(\mathbf{x})
+\hat{d}_{y^2\mathbf{i}}w_{d_{y^2}\mathbf{i}}(\mathbf{x})
+\hat{d}_{xy\mathbf{i}}w_{d_{xy}\mathbf{i}}(\mathbf{x})
\end{equation}
with the sum running over all the lattice sites, and the bosonic creation/annihilation operators obey 
$\left[\hat{d}_{\beta\mathbf{i}},\hat{d}_{\beta\mathbf{j}}^\dagger\right]=\delta_{\mathbf{ij}}$, where 
$\beta=\{x^2,y^2,xy\}$ and all remaining commutators vanish.
Following the usual prescription, $\hat \Psi({\bf x})$ and its Hermitian conjugate $\hat \Psi^\dagger({\bf x})$ are inserted into Eq.~(\ref{many-body_Ham}). We then impose the {\it tight-binding approximation} such that the tunneling is restricted to nearest-neighbour sites and interactions to only dominant onsite terms. This yields the final form of the many-body Hamiltonian describing bosons in the $d$ band. For later convenience, we separate it into two different parts, one containing all the processes that involve the $d_{xy}$ orbital, and one part that only include contributions of the $d_{x^2}$ and $d_{y^2}$ orbitals:
\begin{equation}\label{mbham1}
\hat{H}_\mathrm{BH}=\hat H_\mathrm{p}+\hat H_\mathrm{d}.
\end{equation}
Furthermore, the two separate parts can be split up into  
\begin{equation}\label{mbham2}
\begin{array}{l}
\hat{H}_\mathrm{p}=\hat H_{0p}+\hat H_\mathrm{pkin}+\hat H_\mathrm{pden}+\hat H_\mathrm{pc}+\hat H_\mathrm{po},\\ \\
\hat{H}_\mathrm{d}=\hat H_{0d}+\hat H_\mathrm{dkin}+\hat H_\mathrm{dden}+\hat H_\mathrm{dc}+\hat H_\mathrm{do}.
\end{array}
\end{equation}
The free part of the Hamiltonian accounts for the onsite energies $E_\beta$ of different bands
together with the chemical potential $\mu$ and is given by
\begin{equation}\label{onsen}
H_{0d}=\sum_\beta (E_\beta-\mu)\hat n_{\beta{\bf i}},
\end{equation}
with the particle number operators $\hat n_{\beta{\bf i}}=\hat{d}_{\beta{\bf i}}^\dagger \hat{d}_{\beta{\bf i}}$.
The kinetic energy contributions are
\begin{equation}\label{kinham}
\begin{array}{l}
\displaystyle{\hat H_\mathrm{pkin}  =  -t^\mathrm{p}\sum_{\langle{\bf ij}\rangle}\hat d_{xy{\bf i}}^\dagger\hat d_{xy{\bf j}}},\\ \\
\displaystyle{\hat H_\mathrm{dkin}  =  -\sum_{\alpha, \sigma}\sum_{\langle{\bf ij}\rangle_\sigma}t_\sigma^\alpha
\hat d_{\alpha{\bf i}}^\dagger\hat d_{\alpha{\bf j}}},
\end{array}
\end{equation}
with $\sum_{\langle \bm i,\bm j\rangle_\sigma}$ the sum over nearest neighbours in the direction $\sigma = \{x, y\}$, $\sum_{\langle \bm i,\bm j\rangle}$ the sum over nearest neighbours in all directions 
and $\alpha=\{x^2,\,y^2\}$ the label of the orbital state. The tunneling coefficients satisfy
\begin{equation}
|t_\parallel^\mathrm{\alpha}|>|t^\mathrm{p}|>|t_\perp^\alpha|,
\end{equation}
where $t^\alpha_\parallel$ and $t^\alpha_\perp$ are the tunneling coefficients in the directions parallel and perpendicular to the two nodes of the orbital state. Furthermore $t^\alpha_\parallel,\,t^\alpha_\perp>0$ and $t^\mathrm{p}<0$. Explicit expression of the overlap integrals are given in Appendix~\ref{sec:overlap}. 

The interacting part of the Hamiltonian~(\ref{mbham2}) is characterised by various processes. It contains the 
density-density interactions
\begin{equation}\label{densint}
\begin{array}{l}
\hat H_\mathrm{pden}\!=\!\displaystyle{\sum_{\bf i}\!\left[\frac{U_{p}}{2}\hat n_{xy{\bf i}}(\hat n_{xy{\bf i}}\!-\!1)\!+\!2U_{px}\!\left(\hat n_{x^2{\bf i}}\hat n_{xy{\bf i}}\!+\!\hat n_{y^2{\bf i}}\hat n_{xy{\bf i}}\right)\!\right]\!,}\\ \\
\begin{array}{lll}
\hat H_\mathrm{dden} & = &\displaystyle{\sum_{\bf i}\Bigg\{\frac{U}{2}\left[\hat n_{x^2{\bf i}}(\hat n_{x^2{\bf i}}-1)+\hat n_{y^2{\bf i}}(\hat n_{y^2{\bf i}}-1)\right]}\\ \\
& & +2U_{xy}\hat n_{x^2{\bf i}}\hat n_{y^2{\bf i}}\Bigg\};
\end{array}
\end{array}
\end{equation}
the orbital changing interactions
\begin{equation}\label{flavch}
\begin{array}{lll}
\hat H_\mathrm{pc} & = & \displaystyle{\frac{U_{px}}{2}\sum_{\bf i}\left[\left(\hat d_{xy{\bf i}}^\dagger\hat d_{xy{\bf i}}^\dagger\hat d_{x^2{\bf i}}\hat d_{x^2{\bf i}}+\hat d_{x^2{\bf i}}^\dagger\hat d_{x^2{\bf i}}^\dagger\hat d_{xy{\bf i}}\hat d_{xy{\bf i}}\right)\right.}\\ \\
& & \displaystyle{+\left(\hat d_{xy{\bf i}}^\dagger\hat d_{xy{\bf i}}^\dagger\hat d_{y^2{\bf i}}\hat d_{y^2{\bf i}}+\hat d_{y^2{\bf i}}^\dagger\hat d_{y^2{\bf i}}^\dagger\hat d_{xy{\bf i}}\hat d_{xy{\bf i}}\right)},\\ \\
& & \displaystyle{+\left.\left(\hat d_{xy{\bf i}}^\dagger\hat d_{xy{\bf i}}^\dagger\hat d_{x^2{\bf i}}\hat d_{y^2{\bf i}}+\hat d_{x^2{\bf i}}^\dagger\hat d_{y^2{\bf i}}^\dagger\hat d_{xy{\bf i}}\hat d_{xy{\bf i}}\right)\right]},\\ \\
\hat H_\mathrm{dc} & = & \displaystyle{\frac{U_{xy}}{2}\sum_{\bf i}\left[\left(\hat d_{x^2{\bf i}}^\dagger\hat d_{x^2{\bf i}}^\dagger\hat d_{y^2{\bf i}}\hat d_{y^2{\bf i}}+\hat d_{y^2{\bf i}}^\dagger\hat d_{y^2{\bf i}}^\dagger\hat d_{x^2{\bf i}}\hat d_{x^2{\bf i}}\right)\right]};
\end{array}
\end{equation}
and finally, the density assisted orbital changing collisions
\begin{equation}\label{densass}
\begin{array}{l}
\hat H_\mathrm{po}=\displaystyle{2U_{pxy}\sum_{\bf i}\left(\hat d_{x^2{\bf i}}^\dagger\hat n_{xy{\bf i}}\hat d_{y^2{\bf i}}+\hat d_{y^2{\bf i}}^\dagger\hat n_{xy{\bf i}}\hat d_{x^2{\bf i}}\right)}
\end{array}
\end{equation}
and
\begin{eqnarray}\label{densassdo}
\hat H_\mathrm{do}&=&U_{xxy}\sum_{\bf i}\left[\hat d_{x^2{\bf i}}^\dagger\left(\hat n_{x^2{\bf i}}+\hat n_{y^2{\bf i}}\right)\hat d_{y^2{\bf i}}\right.\\
&+&\left.\hat d_{y^2{\bf i}}^\dagger\left(\hat n_{x^2{\bf i}}+\hat n_{y^2{\bf i}}\right)\hat d_{x^2{\bf i}}\right].
\end{eqnarray}
Other interaction terms vanish due to the symmetries of the orbitals. Like for the tunnelling coefficients, the various coupling constants can again be found in the Appendix~\ref{sec:overlap}. 

As pointed out above, the system of bosons in the $p$ band supports a $\mathbb{Z}_2$-symmetry which can be spontaneously broken to give a vortex-checkerboard condensed phase or a {\it spin-flop phase} in the superfluid or insulating phase respectively. This symmetry arises in that case because $p$-orbital bosons in different orbital states scatter in pairs and therefore the number of particles in each orbital state is preserved modulo 2. On the $d$ band, this conserved quantity is broken by the density-assisted interactions~(\ref{densass}). There exists, however, another non-trivial discrete $\mathbb{Z}_2$-symmetry in the many-body Hamiltonian~(\ref{mbham1}). With the notation $\mathbf{i}=(i_x,i_y)$, we let ${\bf i}'=(i_y,i_x)$, and the Hamiltonian is invariant under the transformation
\begin{equation}\label{symeq}
\begin{array}{l}
\hat d_{x^2{\bf i}}\rightarrow\hat d_{y^2{\bf i}'},\\ \\
\hat d_{y^2{\bf i}}\rightarrow\hat d_{x^2{\bf i}'},\\ \\
\hat d_{xy{\bf i}}\rightarrow\hat d_{xy{\bf i}'}.
\end{array}
\end{equation}
This swaps both the $d_{x^2}$ and $d_{y^2}$ orbitals as well as the lattice indices $i_x$ and $i_y$. Note that the interchange of indices is necessary due to the tunneling anisotropy of the $d_{x^2}$- and $d_{y^2}$-orbital atoms ($t_\sigma^\mathrm{d}\neq t_\sigma^\mathrm{s}$). In the next section we will see that this symmetry can indeed be spontaneously broken to give interesting phases. We should also mention that in addition there are two more $\mathbb{Z}_2$-symmetries. Since the Hamiltonian only consists of quadratic or quartic terms, it is invariant under a sign-change of all operators. This parity symmetry can be further decided into two discrete symmetries, one which change the signs of the $d_{x^2}$ and $d_{y^2}$ orbitals, leaving the $d_{xy}$ orbital unaltered, and one with the reverse operation. These last two symmetries are trivial in the case when the $d_{xy}$ orbital is unpopulated as in Subsecs.~\ref{on1} and \ref{mottsec}.

\subsection{Mean-field approaches}
In the simplest mean-field analysis, the ground-state $|\Psi_0\rangle$  of the many-body system can be expressed as the direct product~\cite{mfref,sachdev}  
\begin{equation}\label{mfstate1}
|\Psi\rangle_\mathrm{MF}=\bigotimes_{\bf i}|\psi_{d_{x^2}{\bf i}},\psi_{d_{y^2}{\bf i}},\psi_{d_{xy}{\bf i}}\rangle_{\bf i},
\end{equation}
of coherent states; $\hat d_{\alpha{\bf i}}|\psi_{d_{x^2}{\bf i}},\psi_{d_{y^2}{\bf i}},\psi_{d_{xy}{\bf i}}\rangle_{\bf i}=
\psi_{d_\alpha{\bf i}}|\psi_{d_{x^2}{\bf i}},\psi_{d_{y^2}{\bf i}},\psi_{d_{xy}{\bf i}}\rangle_{\bf i}$ and
$\hat d_{xy{\bf i}}|\psi_{d_{x^2}{\bf i}},\psi_{d_{y^2}{\bf i}},\psi_{d_{xy}{\bf i}}\rangle_{\bf i}=
\psi_{d_{xy}{\bf i}}|\psi_{d_{x^2}{\bf i}},\psi_{d_{y^2}{\bf i}},\psi_{d_{xy}{\bf i}}\rangle_{\bf i}$.
The classical energy functional is given by $E_\mathrm{cl}\left[\psi_{d_{x^2}{\bf i}},\psi_{d_{y^2}{\bf i}},\psi_{d_{xy}{\bf i}}\right]=\;_\mathrm{MF}\langle\Psi|\hat H_\mathrm{BH}|\Psi\rangle_\mathrm{MF}$ 
and the mean-field solution $|\Psi\rangle_\mathrm{MF}$ is obtained from self-consistently minimising $E_\mathrm{cl}\left[\psi_{d_{x^2}{\bf i}},\psi_{d_{y^2}{\bf i}},\psi_{d_{xy}{\bf i}}\right]$ with respect to the (complex) amplitudes $\psi_{d_\alpha{\bf i}}$ and $\psi_{d_{xy}{\bf i}}$
and the particle conservation constrain 
$\sum_\mathbf{i}\sum_\alpha|\psi_{d_\alpha{\bf i}}|^2+\sum_\mathbf{i}|\psi_{d_{xy}{\bf i}}|^2=N_\mathrm{tot}$, where $N_\mathrm{tot}$ is the total number of particles. 
Note, in particular, that since the Hamiltonian $\hat H_\mathrm{BH}$ is normally ordered~\cite{com1}, the energy expectation values are simply obtained by replacing operators $\hat d_{\alpha{\bf i}}$,
$\hat d_{xy{\bf i}}$, $\hat d_{\alpha{\bf i}}^\dagger$, and $\hat d_{xy{\bf i}}^\dagger$
by the corresponding coherent state amplitudes $\psi_{d_\alpha{\bf i}}$,  $\psi_{d_{xy}{\bf i}}$,
$\psi_{d_\alpha{\bf i}}^*$, and $\psi_{d_{xy}{\bf i}}^*$
 respectively. This crude mean-field method is capable of giving a qualitative correct picture deep in the superfluid phase with a large or moderate number of particles per site, as will be utilised in 
Subsection~\ref{sec:onsite}. However, once the interaction is increased relative to the tunneling, the system becomes insulating and the above approach fails to predict such a phase: the method calls for improvement.

As a next step to advance on this {\em coherent-state ansatz}, (\ref{mfstate1}) we consider intra-site particle fluctuations, but still without intersite correlations
\begin{equation}\label{mfstate2}
|\Psi\rangle_\mathrm{Gutz}=\bigotimes_{\bf i}|\phi_{{\bf i}}\rangle_{\bf i}, 
\end{equation}
where every single site state is 
\begin{equation}\label{mfstate3}
|\phi_{\bf i}\rangle_{\bf i}=\sum_{n_{x^2},n_{y^2},n_{xy}}c_{n_{x^2}n_{y^2}n_{xy}}^{({\bf i})}|n_{x^2},n_{y^2},n_{xy}\rangle_{\bf i}
\end{equation}
with $|n_{x^2},n_{y^2},n_{xy}\rangle_{\bf i}$ a Fock state with $n_{x^2}$ $d_{x^2}$-orbital atoms in site ${\bf i}$ and so on. Note that normalisation implies $\sum_{n_x,n_y,n_{xy}}\left|c_{n_xn_yn_{xy}}^{({\bf i})}\right|^2=1$ for every site ${\bf i}$. The corresponding semi-classical energy functional is similar to the above, $E_\mathrm{Gutz}\left[c_{n_{x^2}n_{y^2}n_{xy}}^{({\bf i})}\right]=\, _\mathrm{Gutz}\langle\Psi|\hat H_\mathrm{BH}|\Psi\rangle_\mathrm{Gutz}$, and the amplitudes $c_{n_{x^2}n_{y^2}n_{xy}}^{({\bf i})}$ are obtained again via self-consistently minimisation of the energy functional.
This procedure, called the {\it Gutzwiller mean-field method}, is particular for its ability of identifying the existence of Mott insulating phases~\cite{BH,gutzref}. However, although the 
factorisation~(\ref{mfstate2}) becomes a more accurate approximation in higher dimensions, it already provides a good estimate of the insulating--superfluid phase transition in 2D~\cite{gutz2d}. 

\section{Phase diagram}
In this section we characterise different phases that can be expected for bosonic atoms in the $d$ bands. To quantitatively determine the phase boundaries of the insulating phases we use the Gutzwiller method discussed in the previous section. For the $p$-band system~\cite{isacsson}, the properties of 
the superfluid phase can be understood from analysing the single site problem on 
a mean-field level and then considering how the presence non-zero tunneling terms establish global long range order. 
Applying the same recipe for the $d$-band system fails for low atomic densities; it suggests that all three orbitals are populated while from the Gutzwiller analysis it follows 
that almost all population is distributed in the 
$d_{x^2}$ and $d_{y^2}$ orbitals. For this reason, when identifying the superfluid state in cases where the number of atoms per site $n_0\lesssim5$ with the coherent state mean-field approach, it is reasonable to constrain the 
analysis to account for only two of the orbitals. At higher densities however, the analysis must include all three orbitals. 

Since mean-field methods are less reliable for probing the physics in Mott insulator phases, this regime will be studied here 
in terms of an effective spin model derived to describe the limit where 
the magnitude of any tunneling coefficient $|t|\ll |U|$, where $U$ is the typical interaction strength.

\subsection{Superfluid-insulator boundaries}
The boundaries separating the Mott insulators from the condensed superfluid phase are determined, as mentioned above, by the Gutzwiller ansatz wave function (\ref{mfstate2}). The condensate order parameters for the three orbitals are given by $\psi_{d_\beta}=_{\,\,\mathrm{Gutz}}\!\!\langle\Psi|\hat d_{d_\beta}|\Psi\rangle_\mathrm{Gutz}$. Within the Gutzwiller method, the insulating phase is described by vanishing order parameters, while it is non-zero in the condensed superfluid phase. 

Before discussing the insulating-superfluid transition we make a brief detour on other possible phases absent in our case due to the imposed approximations. When the lattice amplitude is not too large, the larger width of the higher band Wannier functions may lead to non-negligible contributions from interaction between neighbouring sites. The effective model then includes nearest neighbour interaction terms, which if large enough (compared to the onsite interaction terms), may lead to {\it supersolid phases}~\cite{ss}. The existence of such a phase have been demonstrated for bosonic atoms in the $p$ band~\cite{supersolid} and occurs when the strength of interactions on neighbouring sites is as large as one fourth of the onsite interaction strength. For $30\leq V\leq70$ as considered in this work, however, the strength of interactions on neighbouring sites is always less than $1\%$ and such terms can be safely neglected. Nevertheless, it could be interesting to analyse the possibility of using the $d$ rather than the $p$ bands for realising such novel phases. As one would be required to relax the tight-binding, and most likely also the single-band approximation~\cite{sonia}, this study goes beyond the scope of the present work.

\begin{figure}[h]
\centerline{\includegraphics[width=8cm]{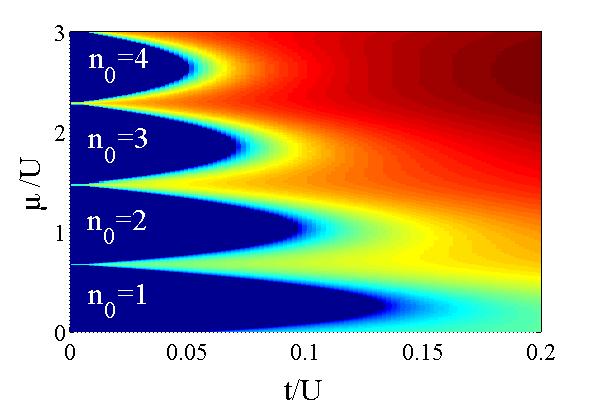}}
\caption{The order parameter $\psi=|\langle\hat{d}_{x^2\mathbf{i}}\rangle|=|\langle\hat{d}_{y^2\mathbf{i}}\rangle|$ (The corresponding order parameter for the $d_{xy}$-orbital is approximately zero except for $t/U<10^{-3}$ where it is, however, at least one order of magnitude smaller than $\psi$). As is seen, the chemical potential is varied such that the first four Mott insulators are illustrated (dark blue regions representing a vanishing superfluid order parameter). For this plot, the relative strengths between the interaction terms has been taken as $\{U_{xy},\,U_p,\,U_{px},\,U_{pxy},\,U_{xxy}\}/U=\{0.17,\,0.9,\,0.3,\,0.04,\,-0.03\}$
corresponding to a lattice with amplitude $V=40$. With this lattice depth, $E_{x^2}=E_{y^2}=0$ and $E_{xy}=1.6U$, and the relative tunnelings $\{t_\perp^\alpha,\,|t^p|\}/t_\parallel^\alpha=\{0.002,\,0.007\}$.} 
\label{fig4}
\end{figure}

Let us now turn to the Gutzwiller predictions for the transitions. The self-consistent minimisation of the Gutzwiller energy functional $E_\mathrm{Gutz}\left[c_{n_{x^2}n_{y^2}n_{xy}}^{({\bf i})}\right]$ is based on the use of the {\it Nelder-Mead simplex algorithm}~\cite{nmalg}. Since we are interested in the phases for the system with different densities, the chemical potential $\mu$ (independent of the orbital flavour) is added to the Hamiltonian, see Eq. (\ref{onsen}). For the numerics we truncate the number of Fock states in the sum (\ref{mfstate3}) to four for every orbital, i.e. up to three atoms for each orbital state and nine atoms in total per site. This gives an onsite Hilbert space dimension of 64, and we can accurately capture the first four Mott lobes. Going to higher insulating states would require Fock states of larger particle number, and thereby a rapid slow down of the numerics. Furthermore, we scale Eq.~(\ref{mbham2}) with the interaction strength $U$.

The results of the numerical Gutzwiller calculation are presented in Fig.~\ref{fig4}, showing the superfluid order parameter in the $\mu t$-plane. In this range of the system's parameters, the occupation of the $d_{xy}$ orbital $\langle \hat n_{xy}\rangle \approx 0$ everywhere. Throughout, proper Wannier functions are used in the computation of the overlap integrals (\ref{coefs}) presented in the Appendix~\ref{sec:overlap}, and we consider the potential amplitudes $30\leq V\leq 70$. These large potential amplitudes (up to 70 recoil energies) are required to ensure the validity of the single-band and tight-binding approximations in the $d$ band, that accommodates very broad Wannier functions.
The order parameters for the $d_{x^2}$ and $d_{y^2}$ orbital are identical and for simplicity we call it $\psi=|\langle\hat{d}_{{x^2}\mathbf{i}}\rangle|=|\langle\hat{d}_{{y^2}\mathbf{i}}\rangle|$. Note that the effective Gutzwiller tunneling $t$ used in the figure is twice the sum of the $s$- and $d$-band tunnelings. More precisely, out of the four nearest-neighbours in the $2D$ system, tunneling processes in the $d$ band occur with (1D) $d$-band tunneling strength to two of the neighbouring sites, and with (1D) $s$-band tunneling strength on the remaining two neighbouring sites. The general structure of the Mott lobes are similar to those found for the regular one orbital BH model using mean-field methods. That is, the extent of the higher lobes fall off as roughly $n_0^{-1}$ for the filling factor $n_0$~\cite{gutz2d}. 

For higher densities $n_0$, the typical interaction energy becomes considerably larger than the energy gap separating the $E_{xy}$ from the  $E_{x^2}$ and $E_{y^2}$ bands, and therefore the occupation of the $d_{xy}$ orbital is expected to increase. This regime can be explored numerically but requires a higher truncation in the number of particles, which makes computations much more costly. To circumvent this issue, we studied the system at fixed chemical potential $\mu/U$ and for various values of $t/U$ and indeed, the results show that with the the same parameters as in Fig.~\ref{fig4}, but with for example $\{\mu,\,t\}/U=\{5,\,0.5\}$ the population of the three orbitals are $(\langle\hat n_{x^2}\rangle,\langle\hat n_{y^2}\rangle,\langle\hat n_{xy}\rangle)\approx(3.3,\,3.3,\,1.7)$.

\subsection{Onsite superfluid states}\label{sec:onsite}
The Gutzwiller mean-field method of the previous Subsection provided us with an estimate of the Mott lobes, but the orbital dependence within the superfluid phase was left implicit. We will approach this
problem next using of the coherent state mean-field ansatz for clarity. 

\subsubsection{Moderate atomic densities}\label{on1}
Generalising the results from the $p$ bands~\cite{isacsson,larson2} we would expect a doubly quantised vortex on every site for the $d$-band system. 
Such a state takes the form
\begin{equation}\label{dvortex}
\begin{array}{c}
\Psi({\bf x})=\sqrt{N_s}/2\left[w_{d_{x^2}}({\bf x})-w_{d_{y^2}}({\bf x})\pm\sqrt{2}iw_{d_{xy}}({\bf x})\right] \\ \\\Leftrightarrow\\ \\
\Psi=\sqrt{N_s}/2\left[\begin{array}{c}
1\\ -1\\ \pm\sqrt{2}i \end{array}\right],
\end{array}
\end{equation}
where $N_s$ is the number of atoms at the site and is fixed {\it a priori}. However, according to the previous Subsection, the occurrence of a state as that of Eq.~(\ref{dvortex}) is precluded for lower atomic densities by the negligible occupation of the 
$d_{xy}$ orbital.  
Moreover, from only $w_{d_{x^2}}({\bf x})$ and $w_{d_{y^2}}({\bf x})$ it is not possible to construct a doubly quantised vortex state, which leads to the conclusion that the superfluid order parameter is a state of a different type.

\begin{figure}[h] 
\centerline{\includegraphics[width=8cm]{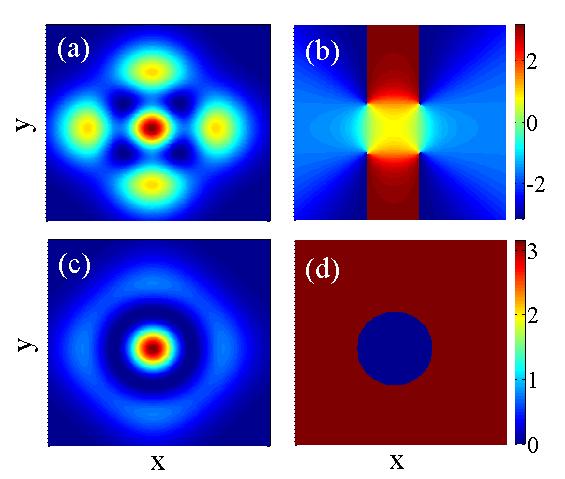}}
\caption{The onsite density $|\Psi({\bf x})|^2/N_s$ (a) and (c) and the corresponding phases $\mathrm{Arg}[\Psi({\bf x})]$ (b) and (d). Here we assumed that the atomic density is rather low such that we can neglect any population of the $d_{xy}$ orbital. In the upper two plots we consider a lattice with an amplitude $V=40$. It is seen in (b) that the phase of the order parameter winds $4\pi$ at the four points where the density vanish. This reflects the presence of four vortices - two vortex/anti-vortex pairs. In the lower two plots we use the same parameters but put $U_{xxy}=-2U_{xy}$ in order to reach the regime where the state qualitatively changes. Here a dark soliton is separating the central peak from the surrounding circle.  } \label{fig3}
\end{figure}

Following the results of the previous Subsection and focusing first on the low-density case, we approximate  
$\psi_{d_{xy}}=0$, and rewrite the single-site order parameter as
\begin{equation}\label{ssstate1}
\Psi=\left[
\begin{array}{c}
\psi_{d_{x^2}}\\ \psi_{d_{y^2}}
\end{array}\right]=\sqrt{N_s}\left[
\begin{array}{c}
\cos\theta \,e^{i\varphi} \\ \sin\theta
\end{array}\right],
\end{equation}
where $0\leq\theta,\leq\pi$ and $0\leq\varphi\,\leq2\pi$. 
Even though this is valid for systems with onsite particle number limited to 10, we believe this gives a good picture of the superfluid state. Indeed, as we show in the next Subsection, a similar state is found when computation includes the third orbital state. 
With the parametrisation of Eq.~(\ref{ssstate1}), the energy functional becomes
\begin{equation}\label{onen}
\begin{array}{lll}
\frac{E_\mathrm{cl}\left[\theta,\varphi\right]}{N_s^2} & = & \displaystyle{\frac{U}{16}\cos4\theta+\frac{U_{xy}}{2}\sin^22\theta\left(\frac{1}{2}+\cos^2\varphi\right)}\\ \\
& & +U_{xxy}\sin2\theta\cos\varphi,
\end{array}
\end{equation}
which after minimisation yields the fixed point  $(\theta_0,\varphi_0)$
\begin{equation}\label{onsol}
\theta_0=\pi/4,\hspace{1cm}\varphi_0=\arccos\left(-U_{xxy}/U_{xy}\right),
\end{equation}
where $\partial_\theta E_\mathrm{cl}[\theta_0,\varphi_0]=\partial_\varphi E_\mathrm{cl}[\theta_0,\varphi_0]=0$. 
This solution,
\begin{equation}
\Psi_\mathrm{vor}({\bf x})=\sqrt{\frac{N_s}{2}}\left[e^{i\varphi_0}w_{d_{x^2}}({\bf x})+w_{d_{y^2}}({\bf x})\right],
\end{equation}
is depicted in Fig.~\ref{fig3} (a) and (b), for the parameters corresponding to a lattice with $V=40$. 
The first plot (a) shows the atomic onsite density $|\Psi_\mathrm{vor}({\bf x})|^2$ and the second (b) the phase $\varphi$ of the order parameter. As is evident, the onsite condensate hosts two vortex/anti-vortex pairs.  Each vortex has a phase winding of
$\pm2\pi$ around each singular point. As argued above, while the na\"ive harmonic approximation on the $d$ bands suggests a single multiply quantised vortex in each site,
they do not occur here due to broken degeneracy of the orbitals. It has also been known
that doubly quantised vortices should be energetically unstable
in a harmonically trapped Bose-Einstein condensate~\cite{vortmult}.

Throughout, proper Wannier functions are used in the computation of the overlap integrals (\ref{coefs}) presented in the Appendix~\ref{sec:overlap}, and we consider the potential amplitudes $30\leq V\leq 70$. These large potential amplitudes (up to 70 recoil energies) are required to ensure the validity of the single-band and tight-binding approximations in the $d$ band, that accommodates very broad Wannier functions. Typical ratios between the interaction strengths for $V=40$ are presented in the caption of Fig.~\ref{fig4}.

The condensates at neighbouring sites have the same ordering of the vortex pairs since 
$t_\parallel^\alpha,\,t_\perp^\alpha>0$, which supports a $``0"$ or $``2\pi"$ phase locking 
between consecutive sites. Note that this is always the solution in case of occupation of only the 
$d_{x^2}$ and $d_{y^2}$ orbitals, and that it does not depends on $U$.  In addition, the configurations defined by the vortices orientations spontaneously break the 
$\mathbb{Z}_2$-symmetry defined in Eq.~(\ref{symeq}) as well as time-reversal symmetry. For the solution given in Fig.~\ref{fig3} (a) and (b), and starting from the vortex in the upper left corner, the phase winds clockwise/counter-clockwise/clockwise/counter-clockwise. The alternative configuration with broken $\mathbb{Z}_2$-symmetry has instead
counter-clockwise/clockwise/counter-clockwise/clockwise winding. 

Now the natural question is what are the properties of the phase with restored $\mathbb{Z}_2$-symmetry? To answer this, we first notice that according to (\ref{onsol}), the shape of the two-vortex pairs depend on the ratio $R\equiv U_{xxy}/U_{xy}$. When $R\rightarrow1$ the vortex/anti-vortex approach each other and annihilates at $R=1$. For $R>1$, the order parameter reads instead
\begin{equation}
\Psi_\mathrm{sol}({\bf x})=\sqrt{\frac{N_s}{2}}\left[w_{d_{x^2}}({\bf x})\pm w_{d_{y^2}}({\bf x})\right].
\end{equation}
The atomic density and the order parameter phase are shown in Fig.~\ref{fig3} (c) and (d). The density vanishes identically around a circle, where also the phase makes a $\pi$-jump from $\varphi_0=0$ to $\varphi_0=\pi$. Such a behaviour describes a {\it dark soliton}, immobile and confined to the lattice site.   

If the ratio $R$ could be externally controlled, then the system could be driven through a phase transition from a `soliton'-superfluid into a symmetry broken `vortex'-superfluid. In addition, since the derivative of the classical energy is discontinuous at the critical point $R_c =1$, it suggests that the transition is of the first order. This is not so surprising, since the states separated by the critical point are topologically different. How to control $R$ experimentally may be non-trivial, but nevertheless, the present analysis sheds some light on the underlying physics of $d$-band bosons.

We end this Subsection by noticing that on the $p$ band in 3D the orbitals are also three-fold degenerate. In this case an artificial instability appears whenever the harmonic approximation is considered, i.e. by replacing the Wannier functions by the corresponding harmonic oscillator eigenfunctions, which implies that it can become favourable to populate the orbitals unequally~\cite{larson2,tomasz}. On the $d$ band, the depopulation of the $d_{xy}$ orbital does not derive from such a spurious effect, but instead, it appears together with the self-consistent minimisation of the full (lattice) energy functional, that includes both the interacting and tunneling processes simultaneously. This means, furthermore, that the combined (independent) knowledge of onsite properties with additional considerations accounting for effects of the tunneling - as one usually proceeds for studying the system in the $p$ band, is not enough to determine the physics in the $d$ band.

\subsubsection{High atomic densities}\label{on2}

\begin{figure}[h] 
\centerline{\includegraphics[width=8cm]{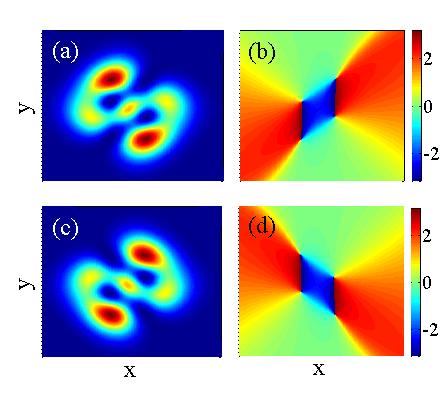}}
\caption{The same as Fig.~\ref{fig3} but deep in the superfluid phase ($n_0\gg1$) where all three orbitals are considerably populated. The difference between the upper and lower plots is the sign of the order parameter $\psi_{d_{xy}}$ (the two states have equal energy), which reflects the states in two neighbouring sites in the lattice. Despite the fact that the $d_{xy}$ orbital is largely populated in this case, the general structure of the superfluid state shown in Fig.~\ref{fig3} (a) and (b) survives, i.e. the onsite order parameter hosts two vortex/anti-vortex pairs. However, the state gets distorted with two of the vortices at the very edge of the distribution. 
As for Fig.~\ref{fig3}, the potential amplitude $V=40$. } \label{fig5}
\end{figure} 

In this Subsection we consider a larger particle number $n_0\gg1$ such that all three orbitals are populated. Relaxing the assumption $\psi_{d_{xy}} = 0$ adds a new variable to be accounted for in the minimisation of the energy functional. By using the parametrisation of the three-orbital order parameter in spherical coordinates, 
\begin{equation}
\begin{array}{l}
\psi_{d_{x^2}}=\sqrt{N_s}\cos\theta\cos\phi e^{i\varphi},\\
\psi_{d_{x^2}}=\sqrt{N_s}\cos\theta\sin\phi e^{i\vartheta},\\
\psi_{d_{xy}}=\sqrt{N_s}\sin\theta,
\end{array}
\end{equation}
the normalization constraint is automatically taken care of and by further choosing the overall phase to be zero we are left with four angle parameters; $E_\mathrm{cl}[\theta,\phi,\varphi,\vartheta]$. 
Since in the general case we do not find analytical solutions for the minimisation (fix point) the system is studied numerically (again utilising the Nelder-Mead simplex algorithm). 

When occupation of the $d_{xy}$-orbital increases, it is in principle possible 
to form doubly excited vortices in each of the lattice sites. However, as argued above, it is questionable whether such state could be energetically favourable. Another aspect arising with all three orbitals populated is that the other two $\mathbb{Z}_2$-symmetries discussed after Eq.~(\ref{symeq}) become relevant as will be demonstrated. These two symmetries are reflected in the fact that the energy $E_\mathrm{cl}$ is invariant under either $(\psi_{d_{x^2}},\psi_{d_{y^2}},\psi_{d_{xy}})
\leftrightarrow(-\psi_{d_{x^2}},-\psi_{d_{y^2}},\psi_{d_{xy}})$ and/or 
$(\psi_{d_{x^2}},\psi_{d_{y^2}},\psi_{d_{xy}})\leftrightarrow(\psi_{d_{x^2}},\psi_{d_{y^2}},-\psi_{d_{xy}})$ (we note that these two are equivalent via a gauge transformation of the overall phase).

The numerically obtained ground state is displayed in Fig.~\ref{fig5} (a) and (b). Here, the population of the three orbitals $d_{x^2}$, $d_{y^2}$, and $d_{xy}$ are 0.36, 0.33, and 0.31 respectively, and hence none of the orbital states dominates the others. The effect of the non-zero population of the $d_{xy}$ orbital is to squeeze and rotate (in this example counter-clockwise) the atomic distribution without destroying the two vortex/anti-vortex pairs; the larger population in the $d_{xy}$ orbital the more rotated and squeezed is the atomic distribution. Thus, as is clear from the deformations in Fig.~\ref{fig5}, vortices of (for example) positive winding are closer to the centre of the site where the density is higher. This suggests that as interactions become stronger and the different orbitals
become effectively more degenerate, the squeezing is also increased and finally two of the vortices appear were the density is vanishingly small. In this case the onsite states start to approach doubly
quantised vortex states. This is an interesting observation, since as we have pointed out earlier such a vortex is typically not stable. However, one should bare in mind that here we have two singly excited vortices coming infinitely close to each other. Furthermore, this continuous deformation does not imply a violation of conserved angular momentum; the state only appears as a doubly excited state, and in the full lattice model the doubly excited vortices would alternate between neighbouring sites such that the total angular momentum vanishes.

To understand the full superfluid state that extends over the whole lattice we recall that the tunneling amplitude $t^p$ on the $p$ band comes with a negative sign. This implies that $\psi_{d_{xy}}$ changes sign between neighbouring sites while $\psi_{d_{x^2}}$ and $\psi_{d_{y^2}}$ do not. This is illustrated in Fig.~\ref{fig5} (c) and (d). The reversed sign of $\psi_{d_{xy}}$ has the effect of rotating the distribution in the opposite direction (i.e. here clockwise), but the winding numbers are unaffected. Now recall that the $\mathbb{Z}_2$-symmetry of Eq.~(\ref{symeq}) corresponds to reversal of the phase winding. Taking all into account we have that in the symmetry broken phase the onsite rotation direction alternates between neighbouring sites (breaking of the parity symmetry $(\psi_{d_{x^2}},\psi_{d_{y^2}},\psi_{d_{xy}})\leftrightarrow(\psi_{d_{x^2}},\psi_{d_{y^2}},-\psi_{d_{xy}})$) and furthermore the phase winding is determined from breaking of the symmetry (\ref{symeq}). In addition, we have numerically checked that in the same way as in the previous subsection, the state in the phase with restored symmetry is characterised by a dark soliton.

\subsection{Insulating phases for filling $n_0=1$}\label{mottsec}
Deep inside the Mott insulating phases the relevant physics of the many-body system can be studied in terms of an effective (pseudo) spin model derived from the perturbative treatment of the tunneling processes relative to the interaction~\cite{spinl}. By mapping the problem onto a spin system, the effective dimension of the Hilbert space is greatly reduced and the relevant physics becomes more transparent and easily analysed. Moreover, even though the properties of many spin models are still unknown, known results of previous studies can be applied to the present case. 

Since the different Mott lobes are characterised by a definite integer filling $n_0$, the derivation of the effective spin model can be most naturally handled with the use of operators that project the eigenvalue problem onto orthogonal particle number subspaces of the Hilbert space. As a consequence, the result of the perturbative treatment is not general to any insulating state, and thus we restrict the study here to the first Mott lobe ($n_0=1$) which is most relevant for experimental studies. We proceed by defining the $\hat P$ and $\hat Q$ operators, $\hat P^2 = \hat P$ and $\hat Q^2 = \hat Q$, that project, respectively, onto the space of states with unit filling, and the states with at least one doubly occupied site. For the lowest Mott insulator we may again impose the assumption that the $d_{xy}$-orbital state is negligibly populated, and that the only relevant degrees of freedom considered stem from the $d_{x^2}$ and $d_{y^2}$ orbitals. 

The eigenvalue problem can then be written as (for a more detailed derivation see the Appendix and Refs.~\cite{fernanda2,fernanda3})
\begin{equation}\label{H_eff}
\hat H_{n_0=1} = -\hat P \hat H_{\text{dkin}}\hat Q\left(\frac{1}{\hat
    Q\hat H_\text{U} \hat Q - E} \right)\hat
Q\hat H_{\text{dkin}}\hat P,
\end{equation}
where $\hat H_{\text{dkin}}$ is given by Eq.~(\ref{kinham}) and the interaction part $\hat H_\text{U}
= \hat H_\text{dden} + \hat H_\text{dc} + \hat H_\text{do}$, cf.
Eqs.~(\ref{densint})-(\ref{densass}). Due to the assumptions of the Mott phase, the resolvent $1/\left(\hat Q\hat H_\text{U} \hat Q - E\right)$ can be expanded such that the effective Hamiltonian contains contributions to  second order in $t/U$. In addition, the tight-binding approximation allows proceeding with the analysis as a two-site problem. Therefore, we define the basis spanning the $\hat P$ subspace by
\begin{equation}
\mathcal{H_P} = \big\{\,\vert x, x\rangle, \vert x, y\rangle, \vert
y, x\rangle, \vert y, y\rangle\big\},
\end{equation}
where $\vert \alpha, \beta\rangle = \hat d^{\dagger}_{\alpha\bf
  i}\hat d^{\dagger}_{\beta\bf j}\vert 0\rangle$ corresponds to the state
with a $d_{\alpha}$-orbital atom in the site $\bf i$ and a
$d_{\beta}$-orbital atom in the neighbouring site $\bf j$, $\alpha, \beta =
\{x, y\}$. In the same way, the relevant states in the basis of the
subspace $\hat Q$ of doubly occupied sites is given by 
\begin{equation}\label{H_Q}
\mathcal{H_Q} = \{\vert 0, 2x\rangle, \vert 0, xy\rangle, \vert 0,
2y\rangle\},
\end{equation}
with $\vert 0, 2\alpha\rangle = 2^{-1/2}\hat d^\dagger_{\alpha \bf
  j}\hat d^\dagger_{\alpha \bf j} \vert 0\rangle$ and $\vert 0, \alpha\beta\rangle
= \hat d^\dagger_{\alpha \bf j} \hat d^\dagger_{\beta,\bf j}\vert 0
\rangle$.

Due to the possibility of transferring population between the different
orbital states via $\hat H_\text{dc}$ and $\hat H_\text{do}$, we
notice, however, that the projection of the Hamiltonian onto the
$\mathcal{H_Q}$ subspace is non-diagonal in the basis of intermediate
states of the perturbation theory. A practical way of dealing
with this situation is to notice that the contributions of the
different processes can be obtained from computation of $(\hat H_Q -
E)$, where $\hat H_Q = \hat Q\hat H_\text{U}\hat Q$, with a subsequent inversion. In addition, since the energy $E \sim t^2/U$ we have  $(\hat H_Q -
E)^{-1}\approx \hat H^{-1}_Q$. More explicitly, ordering the basis of
$\mathcal{H_Q}$ according to~(\ref{H_Q}), 
\begin{equation}
\hat H_Q  = \left(\begin{array}{ccc}
U_{xx} & \sqrt{2}U_{xy} & U_{xy}\\
\sqrt{2}U_{xy} & 2U_{xy} & \sqrt{2}U_{xy}\\
U_{xy} & \sqrt{2}U_{xy} & U_{yy}
\end{array}\right)
\end{equation}
and 
\begin{widetext}
\begin{equation}
\hat H_Q^{-1}  = \Lambda\left(\begin{array}{ccc}
2U_{xy}U_{yy}-U^2_{xy} & \sqrt{2}(U^2_{xy} - U_{xy}U_{yy})  & -2U^2_{xy}\\
\sqrt{2}(U^2_{xy} - U_{xy}U_{yy}) & 2U_{xx}U_{yy} - 2U^2_{xy} &                                           
\sqrt{2}(U^2_{xy} - U_{xy}U_{xx})\\
-2U^2_{xy} & \sqrt{2}(U_{xy}^2 - U_{xy}U_{xx}) & 2U_{xy}U_{xx} - U_{xy}^2,
\end{array}\right)
\end{equation}
\end{widetext}
with 
\begin{equation}\label{lamb}
\Lambda = \left(4U_{xx}U_{yy}U_{xy} - 2U_{xx}U_{xy}^2 - 2U_{yy}U_{xy}^2    \right)^{-1}.
\end{equation}
We determine the final form of the effective Hamiltonian by computing
the relevant matrix elements of Eq.~(\ref{H_eff}). This yields (see
Appendix~\ref{spinoff} for more details of the derivation)
\begin{widetext}
\begin{equation}\label{interm_xyz}
\begin{array}{rcl}
\hat H_{n_0=1}\!\!\!& =\!\!\!&
                                     \displaystyle{-\Lambda\sum_{\alpha\beta}\!\sum_{\langle
                                     \bf i \bf  
                         j\rangle_\sigma}\!\Big\{
                         2|t^\alpha_\sigma|^2\left (2U_{\alpha\alpha}U_{\beta\beta}
                         - U^2_{\alpha\beta}\right)\hat n_{\alpha \bf i}\hat
                         n_{\alpha\bf j}
                                     +2|t^\alpha_\sigma|^2\left(U_{\alpha\alpha}U_{\beta\beta}
                                     - U_{\alpha\beta}^2  
    \right)\hat n_{\alpha \bf i}\hat n_{\beta\bf j}}\\\\
& & \displaystyle{- 2t^\alpha_\sigma
  t^\beta_\sigma U_{\alpha\beta}\hat d^\dagger_{\beta\bf i} \hat
    d_{\alpha\bf i}\hat d_{\beta\bf j}^\dagger \hat d_{\alpha \bf j} + 2t^\alpha_\sigma
    t^\beta_\sigma\left(U_{\alpha\alpha}U_{\beta\beta} -
    U_{\alpha\beta}^2 \right)\hat d^\dagger_{\beta\bf i} \hat
    d_{\alpha\bf i}\hat d_{\alpha\bf j}^\dagger \hat d_{\beta \bf j}}\\\\
& &
    \displaystyle{+\left(|t^\alpha_\sigma|^2 + t^\alpha_\sigma t^\beta_\sigma
    \right)\Big[\left (U_{\alpha\beta}^2 -
    U_{\alpha\beta}U_{\beta\beta}\right)\hat n_{\alpha\bf i}\hat
    d_{\beta\bf 
    j}^\dagger\hat d_{\alpha\bf j} + \left ( U^2_{\alpha\beta}  -
    U_{\alpha\beta}U_{\alpha\alpha}\right)\hat d_{\alpha\bf 
    i}^\dagger\hat d_{\beta\bf i} \hat n_{\alpha\bf j}\Big]\Big\}}
\end{array} 
\end{equation}
\end{widetext}
By further employing the {\it Schwinger angular momentum representation}~\cite{schwing}
\begin{equation}\label{schwing}
\begin{array}{rcl}
\hat S^z_{\bf i} &=& \frac{1}{2}\left(\hat n_{x^2\bf i} - \hat n_{y^2\bf i}\right)\\ \\
\hat S^{+}_{\bf i} &=& \hat S^x_{\bf i} + i\hat S^y_{\bf i}=\hat d_{x^2{\bf i}}^\dagger\hat d_{y^2{\bf i}}\\ \\
\hat S^{-}_{\bf i}&=& \hat S^x_{\bf i} - i\hat S^y_{\bf i}=\hat d_{y^2{\bf i}}^\dagger\hat d_{x^2{\bf i}}
\end{array}
\end{equation}
together with the constraint of unit filling $\hat n_{\bf i}=\hat n_{x^2\bf i} +
\hat n_{y^2\bf i} = 1$, Eq.~(\ref{interm_xyz}) can be mapped into a
spin-$1/2$ $XY\!Z$ model with DM interactions~\cite{dm,dm2} and an
external field 
\begin{equation}\begin{array}{rcl}
\hat H_{XY\!Z} &=&-\displaystyle{\sum_{\langle \bf i
                         \bf j\rangle} J\left[(1 + 
  \gamma)S^x_{\bf i}S^x_{\bf j} + (1-\gamma)S^y_{\bf i}S^y_{\bf j} \right]}\\ \\
& & -\displaystyle{\sum_{\langle \bf i \bf j\rangle}\left[ \Delta S^z_{\bf
  i}S^z_{\bf j}+\delta\left(S^x_{\bf i}S^z_{\bf
    j} + S^z_{\bf i}S^x_{\bf j} \right)\right]+ \sum_{\bf i}\Gamma S^x_{\bf i}}.
\end{array}
\end{equation}
Here $J = t^x_\sigma t^y_\sigma U^2/\Lambda$, with $U^2 =
4(U_{xx}U_{yy} - U_{xy}^2)$ and $\Lambda$ given by
Eq.~(\ref{lamb}), the anisotropy parameter $\gamma = -4U_{xy}^2/U^2$, 
\[
\begin{array}{rcl}
\Delta &=& \displaystyle{|t^x_\sigma|^2\left(2U_{xy}U_{yy} - U_{xy}^2\right) +
|t^y_\sigma|^2\left(2U_{xy}U_{xx} - U_{xy}^2 \right)}\\[0.5em]
& &\displaystyle{ - \left(|t^x_\sigma|^2 + |t^y_\sigma|^2
    \right)\left(U_{xy}^2 + 4U_{xx}U_{yy} \right)},
\end{array}
\]
\[
\begin{array}{rcl}
\delta &=& \left(U_{xy}^2 - U_{xy}U_{yy}
  \right)\left(|t^x_\sigma|^2 + t^x_\sigma t^y_\sigma \right)\\[0.5em]
& &- \left(U_{xy}^2 - U_{xy}U_{xx} \right) \left(|t^y_\sigma|^2 +
  t^x_\sigma t^y_\sigma \right),
\end{array}
\]
and
\[
\begin{array}{rcl}
\Gamma &=& \left(U_{xy}^2 - U_{xy}U_{yy}
  \right)\left(|t^x_\sigma|^2 + t^x_\sigma t^y_\sigma \right)\\[0.5em]
& &+ \left(U_{xy}^2 - U_{xy}U_{xx} \right) \left(|t^y_\sigma|^2 +
  t^x_\sigma t^y_\sigma \right).
\end{array}
\]
However, in the isotropic lattice as considered in this work, the DM
interactions vanish due to the $\mathbb{Z}_2$-symmetry discussed in
Eq.~(\ref{symeq}), which in the spin language read $\hat S_{\bf i}^x\rightarrow\hat S_{\bf i}^x$, $\hat S_{\bf i}^y\rightarrow-\hat S_{\bf i}^y$, and $\hat S_{\bf i}^z\rightarrow-\hat S_{\bf i}^z$. This symmetry is broken in anisotropic lattices. In fact, there 
$t^x_{\parallel} \neq t^y_{\parallel}$ and $t^x_{\perp} \neq
t^y_{\perp}$, and in such case, since $\delta \neq 0$, the DM terms reappear in the effective  spin model.

Several interesting facts should be noted. First, that in the same way
as for the system of bosons in the $p$ band~\cite{fernanda2}, $d$-orbital bosons
provide an alternative realisation of the $XY\!Z$ Heisenberg model,
albeit here with the presence of an external field even in the
isotropic case. This is a consequence of the fact that the density assisted
processes, Eq.~(\ref{densass}), break the $\mathbb{Z}_2$-parity symmetry associated to  
orbital changing interactions of Eq.~(\ref{flavch}) which preserves the number of orbital atoms modulo two. This also explains the presence of the $S^x_{\bm i}$ term in the Hamiltonian and
it readily follows that in the absence of this field the parity symmetry is restored. Second,
that in the same way as for the external field, the DM interaction results from the density-assisted processes. Third, that in the limit
of vanishing density-assisted interactions, the effective spin
Hamiltonian becomes identical to the corresponding one of the $p$-orbital system, of Ref.~\cite{fernanda2}. That this indeed should be
the case can be understood from the fact that the anisotropy parameter $\gamma$ is a consequence of the orbital changing
processes. However, a difference between the two models is that while
the $t^\alpha_\parallel t^\beta_\perp <0$ for the $p$-orbital
system, $t^\alpha_\parallel t^\beta_\perp >0$ in the $d$ band, and
therefore the 
effective spin model favours primarily ferromagnetic alignment at
neighbouring sites in all the pseudo-spin components.

To the best of our knowledge, the phase diagram of this model is not fully known in 
$2D$ even for the simplest case of the isotropic lattice, where $\delta=0$. However, the underlying $\mathbb{Z}_2$-symmetry suggests the possibility of a rich phase diagram. We note that for $\Gamma/J\gg\Delta/J$ ($|\gamma|<1$) the system is characterised by a highly magnetised state in the $x$-spin component, while in the opposite limit it is ferromagnetic in the $z$-component of the spin (noting that $\Delta>0$). The later is a symmetry broken phase with magnetisation in either positive or negative $z$-direction. For $J\sim\gamma,\,\Delta$ another possible symmetry broken ferromagnetic phase could appear with the spin in the $xy$-plane or possibly a gapless {\it floating phase} which exists in the 1D $XY\!Z$ model. In 1D the qualitative features of the $XY\!Z$ phase diagram are known~\cite{sela}. This has 
been studied in terms of $p$-band bosons~\cite{fernanda2}, but following the same procedure of that study to reduce the dimensionality to an effective 1D model one would inevitably generate the DM interaction terms for the system in the $d$ band. This, however, gives interesting possibilities since this model is known to host an interesting phase diagram with ferromagnetic and {\it Luttinger liquid phases}~\cite{xyzpd}.

\section{Concluding remarks}\label{sec:con} 
Motivated by a recent experiment~\cite{dband}, we have studied the zero temperature properties of bosonic atoms on the $d$ bands of an isotropic square optical lattice. 
Due to the particular shapes of the onsite $d$ orbitals, the phases characterising the ground-state of this model are different from those analysed in the past for systems on the $p$ band~\cite{isacsson,liu1,larson2}. This is not surprising if we think of the orbitals in the harmonic approximation where the $p$ orbitals carry angular momentum components $l=\pm1$ and the $d$ orbitals $l=0,\,\pm2$. The $|l|=1$ angular momentum of the $p$ orbital is reflected in the singly excited vortex at each site in the superfluid phase. The higher angular momentum components of the $d$ orbitals implies that the onsite state can take different forms than a state with single vortices. The direct generalisation of the superfluid phase on the $d$ band would be a checkerboard lattice of a single 
$|l|=2$ vortex located at every site. 
However, we found instead that two vortex/anti-vortex pairs appear on each site. The common wisdom is that multiple excited vortices are energetically unstable to form several singly excited vortices~\cite{vortmult}, and one could argue that this explains why we should not find doubly excited vortices in this system. This simplified argument cannot be true here since the total angular momentum should be preserved, which is not the case for a state carrying two vortex/anti-vortex pairs where the total angular momentum vanishes. 

The inter-site phase locking of the different intra-site order parameters is determined by the signs of the different tunneling coefficients. For the $d$ band this implied that the corresponding single site vortex states are ordered in the same way in all the sites. This is in contrast with the checkerboard type (or anti-ferromagnetic) arrangement of vortices in the superfluid phase of the system in the 
$p$ band. This ferromagnetic phase is a symmetry broken phase with respect to the direction of phase windings of the vortices. The spontaneous breaking of the symmetry (\ref{symeq}) is also accompanied by breaking of time-reversal symmetry, which otherwise would imply the possibility of switching between the two degenerate vortex states of different orientation. In the symmetry preserved superfluid state, a dark ring soliton is formed on every site. This phase may, however, not occur naturally in experiments since the strength of the density assisted interaction coefficient must become comparable to the coefficient of the orbital changing interaction. This said, it does not necessarily mean that this phase cannot be monitored but then by external driving like suggested in Ref.~\cite{fernanda2}.

Like for the superfluid phase, the orbital structure also makes the physics of the insulating phases very intriguing. We focused on the lowest filling $n_0=1$, where the Mott phase was shown to feature very interesting properties. With a perturbative treatment, this system was mapped onto an $XY\!Z$ Heisenberg spin model in an external field. The phase diagram of this model is not known in 2D, but by considering limiting cases we argued that the model should posses a very rich phase diagram. We further showed the appearance of DM interactions in the case where the lattice is no longer isotropic. 
In particular, by fine tuning the lattice is is possible to restore the degeneracy between the $d_{x^2}$ and $d_{y^2}$ orbitals and still break the lattice isotropy. In the harmonic approximation this accounts to fulfilling the condition $V_xk_x^2=V_yk_y^2$~\cite{mottp} (in unscaled variables, where $V_\mu$ and $k_\mu$ are the potential amplitudes and wave numbers in the directions $\mu=x,\,y$). Thus, it is possible to study effects arising from the DM interactions, of which the most famous effect is perhaps that of {\it spin canting} where an anti-ferromagnetic state builds a finite magnetisation or where the magnetisation in a ferromagnetic state gets quenched~\cite{dm2}. The presence of these terms imply the breaking of the $\mathbb{Z}_2$-symmetry (\ref{symeq}) and as a result the transitions should turn into first order or of Berezinskii-Kosterlitz-Thouless type~\cite{bkt}. 

In 3D, the $n_0=1$ Mott insulator could be effectively described by an $SU(3)$ pseudo-spin model, as was recently shown for the system in the $p$ band~\cite{fernanda3}. The density assisted orbital chaining collision terms (\ref{densass}) present on the $d$ band would, however, give some additional terms to the effective model in comparison to that emerging on the $p$ band. Now, what could we expect from the physics on other insulating phases?  On the $n_0=2$ insulating phase, the projected (two atoms/site) single site Hilbert space is spanned by the three states $\{|xx\rangle,\,|xy\rangle,\,|yy\rangle\}$. In the Schwinger spin representation~(\ref{schwing}) with the constraint $\hat n_{\bf i}=2$ one would obtain an effective spin-1 $XY\!Z$ model. The integer spin implies the possibility of a 
gapless {\it Haldane phase} to appear~\cite{haldane}. Going up into the Mott lobes with larger filling, the effective spin models would be characterised by higher spin, but as long as we remain in 2D, the (pseudo) spins in the two-orbital case would be generators of the $SU(2)$ group. If occupation of the $\hat d_{xy}$ orbital becomes non-negligible, then the pseudo-spins are the generators of the $SU(3)$ group even in the $2D$ lattice, with a corresponding spin model with DM interactions in all the components.

Since this work presents the first theoretical study of the physics in the $d$ band, we have limited the analysis to the isotropic 2D lattice. Naturally, other lattice geometries and in other dimensions the physics is expected to change as well as if one considers fermionic atoms instead of bosonic ones. Another aspect left out here is the influence of a trapping potential. On the $p$ band it was recently demonstrated that the presence of a harmonic trap is a most simple way to directly detect outcomes of the anisotropic tunneling on the excited bands~\cite{trap}. Similar effects as those found for the confined $p$-band system would also appear on the $d$ band, i.e. {\it time-of-flight detection} of the freely expanding atomic cloud or {\it single-site addressing}~\cite{ssite} would reveal information about the intrinsic anisotropic tunneling on the $d$ bands.

\appendix

\section{Hamiltonian parameters}\label{sec:overlap}
Here we give the general expressions for the overlap integrals rendering the various parameters of the many-body Hamiltonian~(\ref{many-body_Ham}). Making use of the separability of the potential and the Wannier functions~(\ref{wans}), and after introducing the 1D single particle Hamiltonian $\hat H_\mathrm{1sp}=-\partial_x^2+V\sin^2(x)$, the coefficients become
\begin{equation}\label{coefs} 
\begin{array}{l}
\displaystyle{t_\perp^\mathrm{s}=\int dx\,w_{si+1}(x)\hat H_\mathrm{1sp}w_{si}(x)},\\ \\
\displaystyle{t^\mathrm{p}=\int dx\,w_{pi+1}^*(x)\hat H_\mathrm{1sp}w_{pi}(x)},\\ \\
\displaystyle{t_\parallel^\mathrm{d}=\int dx\,w_{di+1}(x)\hat H_\mathrm{1sp}w_{di}(x)},\\ \\
\displaystyle{U=\left(\int dx\,w_{di}^4(x)\right)\left(\int dx\,w_{si}^4(x)\right)},\\ \\
\displaystyle{U_p=\left(\int dx\,|w_{pi}(x)|^4\right)^2},\\ \\
\displaystyle{U_{xy}=\left(\int dx\,w_{di}^2(x)w_{si}^2(x)\right)^2},\\ \\
\displaystyle{U_{xxy}\!=\!\left(\!\int dx\,w_{di}^2(x)w_{si}(x)\!\right)\!\!\left(\!\int dx\,w_{di}(x)w_{si}^3(x)\!\right)\!},\\ \\
\displaystyle{U_{pxy}=\left(\int dx\,|w_{pi}(x)|^2w_{si}(x)w_{di}(x)\right)^2},\\ \\
\displaystyle{U_{px}=\!\left(\!\int dx\,w_{di}^2(x)|w_{pi}(x)|^2\!\right)\!\!\left(\!\int dx\,|w_{si}(x)|^2|w_{pi}(x)|^2\!\right)\!.}
\end{array}
\end{equation}
In the above expressions we have used the phase convention that $w_{si}(x)$ and $w_{di}(x)$ are purely real while $w_{pi}(x)$ is purely imaginary.

\section{Explicit computation of the effective spin model}\label{spinoff}
The final form of the effective Hamiltonian is obtained after
computing the relevant matrix elements of Eq.~(\ref{H_eff}). That is, we
consider all the different transitions allowed for each
state.

The states of the type $\vert \alpha_{\bf i}, \alpha_{\bf j}\rangle$, i.e. the same type of orbital atom in the two neighbouring sites, are connected via tunneling to the three different intermediate states in
the $\mathcal{H_Q}$ subspace; 
\begin{equation*}
\begin{array}{l}
\hat K\, \hat
d^\dagger_{\alpha \bf 
j} \hat d_{\alpha \bf i} \vert\alpha_{\bf i},\alpha_{\bf j}\rangle 
\sqrt{2}\hat K\vert 0,
2\alpha_{\bf j}\rangle\\ \\
= \sqrt{2} \left(
  K^{\alpha\alpha}_{\alpha\alpha}\vert 0, 2\alpha_{\bf j}\rangle +
  K^{\alpha\beta}_{\alpha\alpha}\vert 0, \alpha_{\bf j}\beta_{\bf j}\rangle + 
 K^{\beta\beta}_{\alpha\alpha}\vert 0, 2\beta_{\bf j}\rangle\right),
\end{array}
\end{equation*}
where we have introduced the shorthand notation $\hat K=\hat H_Q^{-1}$. 
The possible transitions are 
\begin{itemize}
\item To $\vert \alpha_{\bf i}, \alpha_{\bf j}\rangle$ via action of
  $\hat d^\dagger_{\alpha\bf i}\hat d_{\alpha\bf j}$, which contribute
  to the effective Hamiltonian with terms of the type
\begin{equation}\label{a2}
\displaystyle{-\sum_{\langle {\bf i} {\bf j}\rangle_\sigma}
\sum_{\alpha, \beta}2\frac{|t^{\alpha}_\sigma|^2}{\Lambda}
\left( U_{\alpha\alpha}
  U_{\beta\beta} - U^2_{\alpha\beta}\right)\hat n_{\alpha
\bf i}\hat n_{\alpha \bf j}}.
\end{equation}

\item To $\vert \alpha_{\bf i},\beta_{\bf j}\rangle $ via action of $\hat
  d^\dagger_{\alpha \bf i}\hat d_{\alpha \bf j}$;
\begin{equation}\label{trans1}
\displaystyle{-\sum_{\langle {\bf i} {\bf j}\rangle_\sigma}
\sum_{\alpha, \beta}2\frac{|t^{\alpha}_\sigma|^2}{\Lambda}
\left( U_{\alpha\alpha}
  U_{\beta\beta} - U^2_{\alpha\beta}\right)\hat n_{\alpha
\bf i}\hat d^\dagger_{\beta \bf j}\hat d_{\alpha \bf j}}.
\end{equation}

\item To $\vert \beta_{\bf i},\alpha_{\bf j}\rangle $ via action of $\hat
  d^\dagger_{\beta \bf i}\hat d_{\beta \bf j}$;
\begin{equation}\label{trans2}
\displaystyle{-\sum_{\langle {\bf i}{\bf j}\rangle_\sigma}
\sum_{\alpha, \beta}2\frac{t^{\alpha}_\sigma t^\beta_\sigma}{\Lambda}
\left( U_{\alpha\alpha}
  U_{\beta\beta} - U^2_{\alpha\beta}\right)\hat
d^\dagger_{\beta \bf i}\hat d_{\alpha \bf i}\hat n_{\alpha 
 \bf j}}.
\end{equation}

\item To $\vert \beta_{\bf i},\beta_{\bf j}\rangle $ via action of $\hat
  d^\dagger_{\beta \bf i}\hat d_{\beta \bf j}$;
\begin{equation}
\displaystyle{\sum_{\langle {\bf i}{\bf j}\rangle_\sigma}
\sum_{\alpha, \beta}2\frac{t^{\alpha}_\sigma t^\beta_\sigma}{\Lambda}
U^2_{\alpha\beta}\,\hat
d^\dagger_{\beta \bf i}\hat d_{\alpha \bf i}\hat d^\dagger_{\beta 
 \bf j}\hat d_{\alpha\bf j}}.
\end{equation}
\end{itemize}
The states of the the type $\vert \alpha_{\bf i}, \beta_{\bf j}\rangle$ are
also connected via tunneling to the three intermediate states in the
$\mathcal{H_Q}$ subspace;
\begin{equation}
\begin{array}{l}
\hat K\hat d_{\alpha\bf i}^\dagger \hat d_{\alpha\bf j}\vert
\alpha_{\bf i}, \beta_{\bf j}\rangle = \hat K\vert 0, \alpha_{\bf
  j}\beta_{\bf j}\rangle = \\ \\\left(K^{\alpha\alpha}_{\alpha\beta}\vert
  0, 2\alpha_{\bf j}\rangle +  K^{\alpha\beta}_{\alpha\beta}\vert 0,
  \alpha_{\bf j}\beta_{\bf j}\rangle +
  K^{\beta\beta}_{\alpha\beta}\vert 0, 2\beta_{\bf j}\rangle\right). 
\end{array}
\end{equation}

Here, in addition to the conjugates of Eqs.~(\ref{trans1})
and~(\ref{trans2}), the other possible transitions are 
\begin{itemize}
\item To $\vert \alpha_{\bf i}, \beta_{\bf j}\rangle$ via action of
  $\hat d^\dagger_{\alpha\bf i}\hat d_{\alpha\bf j}$, which results in the contribution
\begin{equation}
\displaystyle{-\sum_{\langle {\bf i} {\bf j}\rangle_\sigma}
\sum_{\alpha,
  \beta}2\frac{|t^{\alpha}_\sigma|^2}{\Lambda}\left(U_{\alpha\alpha}U_{\beta\beta}
- U^2_{\alpha\beta}\right)\hat n_{\alpha i}\hat n_{\beta j}}.
\end{equation}

\item To $\vert \beta_{\bf i}, \alpha_{\bf j}\rangle$ via action of
  $\hat d^\dagger_{\beta\bf i}\hat d_{\beta\bf j}$; 
\begin{equation}\label{a7}
\displaystyle{-\sum_{\langle {\bf i} {\bf j}\rangle_\sigma}
\sum_{\alpha,
  \beta}2\frac{t^{\alpha}_\sigma
  t^\beta_\sigma}{\Lambda}\left(U_{\alpha\alpha}U_{\beta\beta} 
- U^2_{\alpha\beta}\right)\hat d^\dagger_{\beta i}\hat d_{\alpha \bf
i} \hat d^\dagger_{\alpha\bf j}\hat d_{\beta\bf j}}.
\end{equation}
\end{itemize}
Combining the matrix elements of Eqs.~(\ref{a2})-(\ref{a7}), we derive the effective Hamiltonian~(\ref{interm_xyz}).

\begin{acknowledgements}
The authors acknowledge financial support from VR-Vetenskapsr\aa set (The Swedish Research Council). JL acknowledges KAW (The Knut and Alice Wallenberg foundation).
\end{acknowledgements}


\begin{thebibliography}{999}

\bibitem{rev} M. Lewenstein, A. Sanpera, V. Ahufinger, B. Damski, A. Sen(De), and U. Sen, Adv. Phys. {\bf 56}, 243 (2007); I Bloch, J Dalibard, W Zwerger, Rev. Mod. Phys. {\bf 80}, 885 (2008).

\bibitem{bloch} M. Greiner, O. Mandel, T. Esslinger, T. W. H\"ansch, and I. Bloch, Nature {\bf 415}, 39 (2002); M. Greiner, O. Mandel, T. W. H\"ansch, and I. Bloch, Nature {\bf 419}, 51 (2002).

\bibitem{BH} D. Jaksch, C. Bruder, I. Cirac, P. Zoller, Phys. Rev. Lett. {\bf 81}, 3108 (1998).

\bibitem{singlesitedet}  W. S. Bakr, J. I. Gillen, A. Peng, S. F\"olling, and M. Greiner, Nature {\bf 462}, 74 (2009); J. F. Sherson, C. Weitenberg, M. Endres, M. Cheneau, I. Bloch,	and S. Kuhr, Nature {\bf 467} 68 (2010).

\bibitem{qmagn} J. Struck, C. \"Olschl\"ager, R. Le Targat, P. Soltan-Panahi, A. Eckardt, M. Lewenstein, P. Windpassinger, and K. Sengstock, Science {\bf 333}, 996 (2011); J. Simon, W. S. Bakr, R. Ma, M. E. Tai, P. M. Preiss, and  M. Greiner, Nature {\bf 472}, 307 (2011).

\bibitem{syngauge} M. Aidelsburger, M. Atala, S. Nascimbene, S. Trotzky, Y.-A. Chen, and I. Bloch, Phys. Rev. Lett. {\bf 107}, 255301 (2011); J. Struck, C. \"Olschl\"ager, M. Weinberg, P. Hauke, J. Simonet, A. Eckardt, M. Lewenstein, K. Sengstock, and P. Windpassinger, Phys. Rev. Lett. {\bf 108}, 225304 (2012); M. Aidelsburger, M. Atala, M. Lohse, J. T. Barreiro, B. Paredes, and I. Bloch, Phys. Rev. Lett. {\bf 111}, 185301 (2013).

\bibitem{topo}  G. Jotzu, M. Messer, R. Desbuquois,	M. Lebrat,	T. Uehlinger, D. Greif	, and T. Esslinger, Nature {\bf 515}, 237 (2014).

\bibitem{FH} R. J\"{o}rdens, N. Strohmaier, K. G\"u nter, H. Moritz, and T. Esslinger, Nature {\bf 455}, 204 (2008).

\bibitem{dynamics}  S. Trotzky, Y.-A. Chen, A. Flesch, I. P. McCulloch, U. Schollw\"{o}ck, J. Eisert, and I. Bloch, Nature Phys. {\bf 8}, 325 (2012);  M. Cheneau,	P. Barmettler, D. Poletti, M. Endres, P. Schau\ss, T. Fukuhara,	C. Gross,	I. Bloch, C. Kollath, and S. Kuhr, Nature {\bf 481}, 484 (2012).

\bibitem{orbital} B. Keimer and A. M. Oles, New J. Phys. {\bf 6}, 1088 (2004).

\bibitem{he3} A. J. Leggett, Rev. Mod. Phys. {\bf 47}, 331 (1975).

\bibitem{metal} M. Imada, A. Fujimori, and Y. Tokura, Rev. Mod. Phys. {\bf 70}, 1039 (1998); Y. Tokura and N. Nagaosa, Science {\bf 288}, 462 (2000).

\bibitem{dance} M. Lewenstein and W. V. Liu, Nature Phys. {\bf 7}, 101 (2011).

\bibitem{isacsson} A. Isacsson and S. M. Girvin, Phys. Rev. A {\bf 72}, 053604 (2005).

\bibitem{liu1} W. V. Liu and C. Wu, Phys. Rev. A {\bf 74}, 013607 (2006).

\bibitem{larson2}  A. Collin, J. Larson, and J.-P. Martikainen, Phys. Rev. A {\bf 81}, 023605 (2010).

\bibitem{mottp} X. Li, Z. Zhang, and W. V. Liu, Phys. Rev. Lett. {\bf 108}, 175302 (2012).

\bibitem{fernanda2} F. Pinheiro, G. M. Bruun, J.-P. Martikainen, and J. Larson, Phys. Rev. Lett. {\bf 111}, 205302 (2013).

\bibitem{pbandphases} C. Wu, W. V. Liu, J. Moore, and S. Das Sarma, Phys. Rev. Lett. {\bf 97}, 190406 (2006).

\bibitem{frustration} C. Wu, Phys. Rev. Lett. {\bf 100}, 200406 (2008). 

\bibitem{fernanda3} F. Pinheiro, arXiv:1410.7828.

\bibitem{pbandfermion} C. Wu, Phys. Rev. Lett. {\bf 101}, 186807 (2008); E. Zhao and W. V. Liu, Phys. Rev. Lett. {\bf 100}, 160403 (2008); C. Wu and S. Das Sarma, Phys. Rev. B {\bf 77}, 235107 (2008); K. Sun, W. V. Liu, A. Hemmerich, and S. D. Sarma, Nature Phys. {\bf 8}, 67 (2012);  X. Li, E. Zhao, and W. V. Liu, Nature Commun. {\bf 4}, 1523 (2013).

\bibitem{latticeacc} J. H. Denschlag, J. E. Simsarin, H. H\"affner, C. McKenzie, A. Browaeys, D. Cho, K. Helmerson, S. L. Rolston, and W. D. Phillips, J. Phys. B {\bf 35}, 3095 (2002); A. Browaeys, H. H\"affner, C. McKenzie, S. L. Rolston, K. Helmerson, and W. D. Phillips, Phys Rev. A {\bf 72}, 053605 (2005)

\bibitem{pbandbloch} T. M\"uller, S. F\"olling, A. Widera, and I. Bloch, Phys.
Rev. Lett. {\bf 99}, 200405 (2007).

\bibitem{pbandhemmerich}  G. Wirth, M. \"Olschl\"ager, and A. Hemmerich, Nature {\bf 7}, 147 (2011).

\bibitem{hemmerich2} T. Kock, M. \"Olschl\"ager, A. Ewerbeck, W.-M. Huang, L. Mathey, and A. Hemmerich, arXiv:1411.3483.

\bibitem{fbandhemmerich}  M. \"Olschl\"ager, G. Wirth, and A. Hemmerich, Phys. Rev. Lett.  {\bf 106}, 015302 (2011).

\bibitem{dband} Y. Zhai, X. Yue, Y. Wu, X. Chen, P. Zhang, and X. Zhou, Phys. Rev. A {\bf 87}, 063638 (2013). 

\bibitem{BH1} M. P. A. Fisher, P. B. Weichman, G. Grinstein, and D. S. Fisher, Phys. Rev. B {\bf 40}, 546 (1989). 

\bibitem{sela} E. Sela, A. Altland, and A. Rosch, Phys. Rev. B {\bf 84}, 085114 (2011).

\bibitem{supersolid} V. W. Scarola, E. Demler, and S. Das Sarma, Phys. Rev. A {\bf 73}, 051601(R) (2006).

\bibitem{mfref} C. Pethick and H. Smith, {\it Bose-Einstein Condensation in Dilute Gases}, (Cambridge University Press, Cambridge, 2002).

\bibitem{sachdev} S. Sachdev, {\it Quantum Phase Transitions}, (cambridge University Press, Cambridge, 2011).

\bibitem{com1} To be precise, to normally order the Hamiltonian we need to commute a pair of operators: $\hat n_{\alpha{\bf i}}(\hat n_{\alpha{\bf i}}-1)\rightarrow\hat a_{\alpha{\bf i}}^\dagger\hat a_{\alpha{\bf i}}^\dagger\hat a_{\alpha{\bf i}}\hat a_{\alpha{\bf i}}$.

\bibitem{gutzref} L. Amico and V. Penna, Phys. Rev. Lett. {\bf 80}, 2189 (1998);D. Jaksch, V. Venturi, J. I. Cirac, C. J. Williams, and P. Zoller, Phys. Rev. Lett. {\bf 89}, 040402 (2002).

\bibitem{gutz2d} J. K. Freericks and H. Monien, Europhys. Lett. {\bf 26}, 545 (1994).

\bibitem{ss} K. Goral, L. Santos, and M. Lewenstein, Phys. Rev. Lett. {\bf 88}, 170406 (2002).

\bibitem{sonia} J. Larson, S. Fernandez-Vidal, G. Morigi, and M. Lewenstein, New J. Phys. {\bf 10}, 045002 (2008).

\bibitem{nmalg} J. E. Daniels and D. J. Woods, {\it New computing environments: microcomputers in large-scale computing}, pages 116-122, (Siam, Philadelphia, 1987).

\bibitem{vortmult} D. A. Butts and D. S. Rokhsar, Nature {\bf 397}, 327 (1999); Y. Castin and R. Dum, Eur. Phys. J. D {\bf 7}, 399 (1999); G. M. Kavoulakis, B. Mottelson, and C. J. Pethick, Phys. Rev. A {\bf 62}, 063605 (2000);E. Lundh, Phys. Rev. A {\bf 65}, 043604 (2002).

\bibitem{tomasz} T. Sowinski, M. Lacki, O. Dutta, J. Pietraszewicz, P. Sierant, M, Gajda, J. Zakrzewski, and M. Lewenstein, Phys. Rev. Lett. {\bf 111}, 215302 (2013).
 
\bibitem{spinl} L.-M. Duan, E. Demler, and M. D. Lukin, Phys. Rev. Lett. {\bf 91}, 090402 (2003).
 
\bibitem{schwing}  J. J. Sakurai, {\it Modern quantum mechanics}, (Adison-Wesley, 1993); A. Auerback,
{\it Interacting electrons and quantum magnetism}, (Springer Verlag, New York, 1998).

\bibitem{dm} I. Dzyaloshinsky, Sov. Phys. JEPT {\bf 5}, 1259 (1957).

\bibitem{dm2} T Moriya, Phys. Rev. {\bf 120}, 91 (1960).

\bibitem{xyzpd} S. Peotta, L. Mazza, E. Vicari, M. Polini, R. Fazio, and D. Rossini, J. Stat. Mech. P09005 (2014).

\bibitem{bkt} V. L. Berezinskii, Sov. Phys. JETP {\bf 32}, 493 (1971); J. M. Kosterlitz and D. Thouless, J. Phys. C {\bf 5}, L124 (1972).

\bibitem{haldane} F. D. M. Haldane, Phys. Rev. Lett. {\bf 50}, 1153 (1983).

\bibitem{trap} F. Pinheiro, J.-P. Martikainen, and J. Larson, Phys. Rev. A {\bf 85}, 033638 {2012}; J. Larson, arXiv:1310.7867.

\bibitem{ssite} W. S. Bakr, J. I. Gillen, A. Peng, S. F\"olling, andM. Greiner, Nature {\bf 462}, 74 (2009); C. Weitenberg, M. Endres, J. F. Sherson, M. Cheneau, P. Schauss, T. Fukuhara, I. Bloch, and S. Kuhr, Nature {\bf 471}, 319 (2011).

\end{thebibliography}
\end{document}